\documentclass[preprint,prc,floatfix,showpacs]{revtex4}

\usepackage{graphicx}
\usepackage{dcolumn}
\usepackage{bm}
\begin{document}

\preprint{PRC Draft $^{61}$Cr $\beta$ decay}

\title{Low-energy structure of $^{61}$Mn populated following
$\beta$ decay of $^{61}$Cr}

\author{
H.L.~Crawford,$^{1,2}$
P.F.~Mantica,$^{1,2}$
J.S.~Berryman,$^{1,2}$
R.~Broda,$^{3}$
B.~Fornal,$^{3}$
C.R.~Hoffman,$^{4}$
N.~Hoteling,$^{5,6}$
R.V.F.~Janssens,$^{5}$
S.M.~Lenzi,$^{7,8}$
J.~Pereira,$^{2,9}$
J.B.~Stoker,$^{1,2}$
S.L.~Tabor,$^{4}$
W.B.~Walters,$^{6}$
X.~Wang,$^{5,10}$ and
S.~Zhu$^{5}$
}

\affiliation{$^{1}$
Department of Chemistry, Michigan State University,
East Lansing, Michigan 48824, USA}
\affiliation{$^{2}$ 
National Superconducting Cyclotron
Laboratory, Michigan State University,
East Lansing, Michigan 48824, USA} 
\affiliation{$^{3}$
Institute of Nuclear Physics, Polish Academy of Sciences, 
Cracow PL-31342, Poland}
\affiliation{$^{4}$
Department of Physics and Astronomy, Florida State University,
Tallahassee, FL 32306, USA}
\affiliation{$^{5}$
Physics Division, Argonne National Laboratory, Argonne, Illinois 60429, USA}
\affiliation{$^{6}$
Department of Chemistry and Biochemistry, University of Maryland,
College Park, MD 20742, USA}
\affiliation{$^{7}$
Department of Physics, Padova University, I-35131 Padova, Italy}
\affiliation{$^{8}$
National Institute of Nuclear Physics, Padova Section, I-35131 Padova, Italy}
\affiliation{$^{9}$
Joint Institute for Nuclear Astrophysics, Michigan State University,
East Lansing, MI  48824, USA}
\affiliation{$^{10}$
Department of Physics, University of Notre Dame,
South Bend, IN 46556, USA}

\date{\today}

\begin{abstract}
$\beta$ decay of the $^{61}$Cr$_{37}$ ground state has been studied.  A new 
half-life of $233 \pm  11$~ms has been deduced, and seven delayed 
$\gamma$ rays have been assigned to the daughter, $^{61}$Mn$_{36}$.  The
low-energy level structure of $^{61}$Mn$_{36}$ is similar to
that of the less neutron-rich $^{57,59}$Mn nuclei.  
The odd-$A$ $_{25}$Mn isotopes follow the systematic trend in the yrast states of the 
even-even, $Z+1$ $_{26}$Fe isotopes, and not that of the 
$Z-1$ $_{24}$Cr isotopes, where a possible onset of collectivity has been suggested to
occur already at $N=36$.  
\end{abstract}


\maketitle

\section{Introduction}

According to the shell model, the $1g_{9/2}$ single-particle
orbital is well separated from the $1g_{7/2}$, $2d_{5/2}$
and other higher-energy orbitals, giving rise to the well-established
$N,Z=50$ magic numbers.  A less pronounced subshell closure for
$N,Z=40$ is also expected, as a smaller
energy gap typically occurs between the
$1g_{9/2}$ level and the lower $pf$-shell model states.
However, collectivity
is evident for most nuclei with $N \sim Z \sim 40$, and
rotational-like yrast structures have been
observed in the even-even, $N=Z$
isotopes $^{76}$Sr and $^{80}$Zr \cite{lis1990,lis1987,fis2001}.
The $Z = 40$ and $N = 40$ shell gaps appear to be more 
robust for neutron-rich nuclei.  Both $^{90}$Zr$_{50}$ and 
$^{96}$Zr$_{56}$ exhibit features at low excitation energy 
that support the presence of a subshell gap at 
$Z=40$ \cite{aue1965,glo1975,fed1979}.
This quasi-magic gap at $Z=40$ disappears with
the addition of neutrons beyond $N=56$, and such 
a behavior was attributed to the strong, attractive monopole
interaction between $1g_{9/2}$ protons and $1g_{7/2}$ 
neutrons \cite{fed1977}. 
Evidence for magicity at $N=40$ for neutron-rich nuclei
is gathered mainly from measurements on  
$^{68}$Ni$_{40}$ and adjacent nuclei.
The energy of the first $2^{+}$ state [$E(2^+_1)$] in 
$^{68}$Ni is 2033~keV \cite{bro1995}, much higher than 
the $E(2^+_1)$ values for the even-even neighbors 
$^{66}$Ni$_{38}$ and $^{70}$Ni$_{42}$.
The ratio $E(4^+_1)$/$E(2^+_1)$ for the even-even Ni 
isotopes reaches a local minimum at $^{68}$Ni, another indication
for added stability at $Z=28, N=40$.  The low
transition probability for excitation to the 
first $2^+$ state in $^{68}$Ni \cite{sor2002,bre2008},  
relative to neighboring isotopes, is also in line with a double-magic
character for this nucleus and the ``goodness'' of the  
subshell closure at $N=40$ for $_{28}$Ni.
  
\begin{figure}[h]
\includegraphics[width=0.5\textwidth]{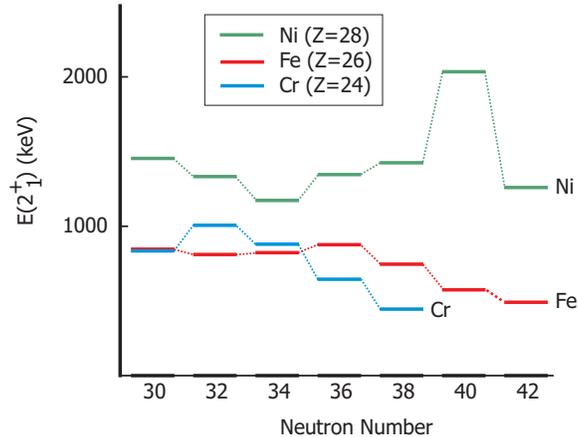}
\caption{(Color online) Systematic variation of $E(2^+_1)$ for the 
even-even $_{28}$Ni, $_{26}$Fe and $_{24}$Cr isotopes.}
\label{fig:systematics}
\end{figure}

However, it seems that the subshell gap at $N=40$ 
may be quickly reduced with the removal or addition of protons.  
As an example, the 
systematic variation of $E(2^+_1)$ energies as a function of neutron number
for the $_{28}$Ni, $_{26}$Fe, and $_{24}$Cr isotopes is presented in 
Fig.\ \ref{fig:systematics}.  The $E(2^+_1)$ values for the 
$_{26}$Fe and $_{24}$Cr isotopes decrease 
with increasing neutron number, even through $N=40$.  
Hannawald {\it et al.} \cite{han1999} identified the first 
$2^+$ state in $^{66}$Fe$_{40}$ from the $\beta$ decay of $^{66}$Mn.  The low 
value $E(2^+_1)$ = 573~keV was taken as an indication 
of possible collectivity near the ground state, and a deformation parameter 
$\beta _{2} = 0.26$ was deduced from the Grodzins relation \cite{gro1962}.
The decreasing trend in $E(2^+_1)$ values for the neutron-rich Fe
isotopes continues in $^{68}$Fe$_{42}$, where the
$2^+_1$ level with energy 517~keV was identified by Adrich {\it et al.} \cite{adr2008}
with in-beam $\gamma$-ray spectroscopy following $2p$ knockout.

The systematic decrease in $E(2^+_1)$ values is even more pronounced 
in the Cr isotopic chain (Fig.\ \ref{fig:systematics}). 
Deformation parameters $\beta _{2} = 0.27$ and  $0.31$ have been deduced 
for the ground states of $^{60}$Cr$_{36}$ and $^{62}$Cr$_{38}$, respectively \cite{sor2003},
based on their $E(2^+_1)$ values.  Large quadrupole deformation
lengths have also been reported for $^{60,62}$Cr based on 
results of  proton inelastic scattering experiments \cite{aoi2008,aoi2009}. 
However, the medium-spin structure of $^{60}$Cr does not display features
characteristic of a rotational pattern \cite{zhu2006}.
Although data for the low-energy structure of $^{64}$Cr$_{40}$ are still lacking,
the trend of a lowering of excitation energies for the neutron-rich
$_{24}$Cr isotopes at and beyond $N=40$ is predicted \cite{kan2008}.

The strength of the subshell closure at $N = 40$ depends
on the magnitude of the gap between the $1f_{5/2}, 2p_{1/2}$ and 
$1g_{9/2}$ neutron orbitals. The strong monopole interaction 
between protons in the $1f_{7/2}$ shell and the neutrons in the $1g_{9/2}$ 
orbital decreases when $1f_{7/2}$ protons are removed, causing the 
narrowing of the $1f_{5/2}$-$1g_{9/2}$ energy separation \cite{han1999}.
In addition, the first two 
Nilsson substates of the $1g_{9/2}$ orbital are steeply downsloping 
with increasing deformation.  Combined, these effects can create a situation 
in which the occupation of the deformed levels is energetically preferential 
to the occupation of the spherical states, leading to increased deformation 
and erosion of the $N=40$ subshell closure for 
$Z < 28$.  The appearance of low-energy $9/2^+$ states 
in the odd-$A$ Cr isotopes gives preliminary evidence that the 
$1g_{9/2}$ neutron orbital is indeed nearing the Fermi surface.  The 
rotational bands built on the $9/2^+$ state in $^{55}$Cr$_{31}$ \cite{app2003}
and $^{57}$Cr$_{33}$ \cite{dea2005} suggest the presence of prolate-deformed structures. 
However, the isomeric nature of the $9/2^+$ level in $^{59}$Cr$_{35}$ \cite{fre2004} appears 
to be more indicative of modest oblate collectivity.     
 
While the neutron-rich, even-$Z$ $_{26}$Fe and $_{24}$Cr nuclei near $N=40$ 
have been investigated, scant data are available regarding the level 
structures of the neighboring odd-$Z$ $_{25}$Mn.  
The driving force leading to the lowering of the $E(2^+_1)$ energies and 
possible onset of collectivity 
in the Cr and Fe isotopes near $N=40$ should be expected to act
in a similar manner in the Mn nuclei.  
However, level structures are not 
available for Mn isotopes with $N > 38$.  
In-beam $\gamma$ rays were recently measured for neutron-rich $^{59-63}$Mn 
produced by multi-nucleon transfer between a $^{70}$Zn projectile
and $^{238}$U target \cite{val2008}.  Only a single $\gamma$-ray transition
was attributed to $^{63}$Mn$_{38}$.  Five $\gamma$ rays were 
assigned to the low-energy structure of $^{62}$Mn$_{37}$, complementing the 
low-energy level scheme that had previously been proposed for 
this nucleus from $\beta$ decay \cite{gau2005}.  The low-energy
structure of $^{62}$Mn is further complicated by the 
presence of two $\beta$-decaying states \cite{gau2005}.
In-beam and $\beta$-delayed $\gamma$ rays have likewise 
been assigned to $^{61}$Mn$_{36}$.  However, no level scheme
was proposed based on the five $\beta$-delayed $\gamma$ rays 
reported by Sorlin {\it et al.} \cite{sor2000}.  
The low-energy structure of $^{61}$Mn offers an important opportunity
to characterize further the possible onset of collectivity inferred
from the systematic behavior of the 
even-even Cr and Fe isotopes.  The decrease in
$E(2^+_1)$ values for neutron-rich Cr may suggest that
deformation sets in at $^{60}$Cr, with 
$N=36$, while evidence for such deformation effects in the 
Fe isotopes appears only at $^{64}$Fe, with $N=38$.

Here, we report on the low-energy structure of $^{61}$Mn$_{36}$, 
populated following the $\beta$ decay of $^{61}$Cr. 
We also examine the systematic variation of the low-energy level
densities of the odd-$A$ $_{25}$Mn isotopes and find no evidence for
an early onset of collectivity at $N =36$, 
as was proposed for neighboring $_{24}$Cr nuclei.

\section{Experimental Procedure}

The $\beta$-decay properties of $^{61}$Cr were studied 
at National Superconducting Cyclotron Laboratory (NSCL) 
at Michigan State University.  A 130-MeV/nucleon 
$^{76}$Ge$^{30+}$ beam was produced by the coupled cyclotrons at NSCL.  
The $^{76}$Ge primary beam was fragmented on a 47 mg/cm$^{2}$ Be 
target at the object position of the A1900 fragment separator 
\cite{mor2003}.  The secondary fragments of interest 
were separated in the A1900 with a 300 mg/cm$^{2}$ Al wedge 
located at the intermediate image of the separator.  
The full momentum acceptance of the A1900 ($\Delta p/p \sim 5$\%) was used for 
fragment collection.

\begin{figure}[h]
\includegraphics[width=0.5\textwidth]{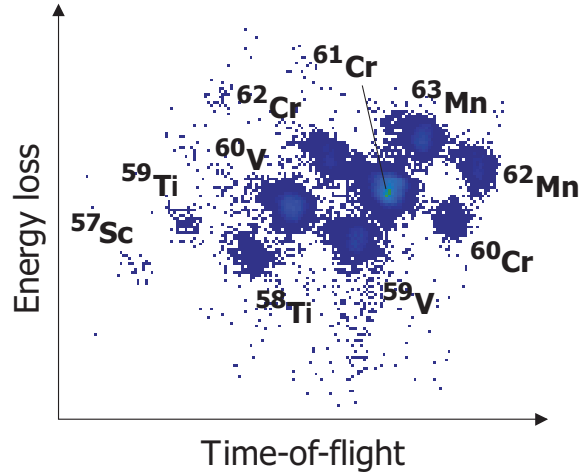}
\caption{(Color online) Particle identification plot measured at the 
experimental end station (BCS) for the A1900 magnetic rigidity 
setting $B\rho _1 = 4.403$~Tm and $B\rho _2 = 4.134$~Tm.
Energy loss was measured in the most upstream PIN detector of the 
BCS.  Time-of-flight was deduced from the difference in 
timing signals of the same PIN detector and a plastic scintillator
at the intermediate image of the A1900.  An event-by-event 
momentum correction of the time-of-flight was performed 
based on the position information obtained from
the plastic scintillator.}
\label{fig:cr61pid}
\end{figure}

Fragments were delivered to the $\beta$-decay experimental end station, 
which consisted of detectors 
from the Beta Counting System (BCS) \cite{pri2003} and the 
Segmented Germanium Array (SeGA) \cite{mue2001}.  
A stack of 3 Si PIN detectors with thicknesses 991, 997 and 309~$\mu$m, 
respectively, was placed upstream of the BCS and provided energy 
loss information for particle identification.  
Fragments were implanted in the 979~$\mu$m-thick double-sided silicon 
microstrip detector (DSSD) of the BCS.
This detector was segmented into 40 strips on both front and 
back, for a total of 1600 pixels.  
A total of $1.32 \times 10^{4}$ $^{61}$Cr ions was 
implanted into the DSSD over the course of the measurement.  
The particle identification spectrum is presented in Fig.\ \ref{fig:cr61pid}.

Delayed $\gamma$ rays were detected by 16 Ge detectors from SeGA, 
arranged in two concentric circles around the BCS detectors.  
The $\gamma$-ray peak detection efficiency varied from 20\% at 100~keV to 
7\% at 1~MeV.  The energy resolution of each Ge detector was 
measured to be better than 3.5~keV for the 1.3-MeV $\gamma$ ray in $^{60}$Co.

\section{Results}

\begin{figure}[h]
\includegraphics[width=0.5\textwidth]{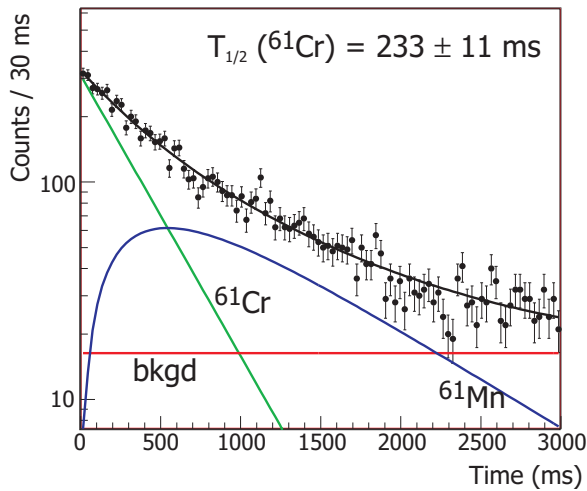}
\caption{(Color online) Decay curve for $^{61}$Cr, based on fragment-$\beta$ 
correlations. Data were fitted with a single exponential decay and exponential 
growth and decay of the short-lived daughter, $^{61}$Mn ($T_{1/2} = 670 \pm 40$~ms).  
A constant background was also considered in the fit.}
\label{fig:halflife}
\end{figure}

Implanted $^{61}$Cr fragments were correlated with 
their subsequent $\beta$ decays by requiring the presence of a 
high-energy implantation event in a single pixel 
of the DSSD, followed by a low-energy $\beta$ event 
in the same or any of the eight neighboring pixels.  
The decay curve for $^{61}$Cr-correlated $\beta$ decays given in 
Fig.\ \ref{fig:halflife} was generated by histogramming 
the differences between absolute time stamps for 
implantation and correlated decay events.  The curve was fitted 
with a single exponential decay combined with an exponential growth and decay 
of the short-lived daughter, $^{61}$Mn, whose half-life was 
taken to be $670\pm 40$~ms as adopted in Ref.\ \cite{bha1999}.  
A constant background was also included as a free parameter in the fit.  
A half-life of $233 \pm 11$~ms was deduced for the 
ground-state $\beta$ decay of $^{61}$Cr.  This new value compares 
favorably with a previous measurement by Sorlin {\it et al.} 
\cite{sor2000} of $251 \pm 22$~ms, but is more than $1\sigma$ shorter 
than the $270 \pm 20$~ms value deduced earlier by 
Ameil {\it et al.} \cite{ame1998}.

\begin{figure}[ht]
\includegraphics[width=0.5\textwidth]{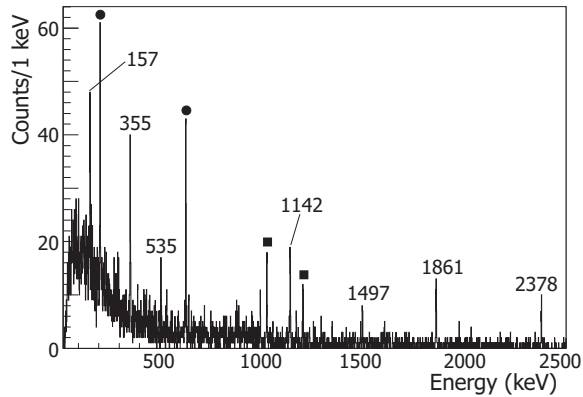}
\caption{$\beta$-delayed $\gamma$ rays following the decay of $^{61}$Cr.  
Known $\gamma$ rays from the decay of the daughter $^{61}$Mn and the 
granddaughter $^{61}$Fe are indicated by the filled circles and 
squares, respectively.  
Transitions assigned to the decay of $^{61}$Cr are marked by their energy in keV.}
\label{fig:gamma_Cronly}
\end{figure}

The $\beta$-delayed $\gamma$-ray spectrum for 
the decay of $^{61}$Cr is shown in Fig.\ \ref{fig:gamma_Cronly}.   
The spectrum covers the energy range 0 to 2.5~MeV, and 
the observed $\gamma$ rays correspond to events 
that occurred within the first 3~s following a $^{61}$Cr implantation.  
Seven transitions have been assigned to the decay of $^{61}$Cr 
and are listed in Table\ \ref{tab:gammas}.
The peaks observed with energies of 207 and 629 keV 
in Fig.\ \ref{fig:gamma_Cronly} are known 
transitions in the decay of the $^{61}$Mn daughter \cite{run1985}.  
The 1028- and 1205-keV transitions in the decay of the 
grand-daughter $^{61}$Fe \cite{bro1975} are also seen in Fig.\ \ref{fig:gamma_Cronly}. 
The peaks at 355, 535, 1142 and 1861~keV correspond in 
energy to the transitions previously observed with 
low statistics by Sorlin {\it et al.} \cite{sor2000}.  
However, the transition observed in Ref.\ \cite{sor2000} at 1134~keV 
was not apparent in the spectrum of Fig.\ \ref{fig:gamma_Cronly}.  
The three additional $\gamma$-ray transitions at 157, 1497, and 2378 keV
are assigned for the first time to the decay of $^{61}$Cr.
Coincidentally, two transitions with energies
155 and 355 keV had previously
been assigned to the $\beta$ decay of $^{62}$Cr \cite{gau2005}. 
Observation of these transitions in Fig.\ \ref{fig:gamma_Cronly}
suggests that they are associated with levels in $^{61}$Mn, and their
observation in Ref.\ \cite{gau2005} may be evidence for 
$\beta$-delayed neutron decay of the $^{62}$Cr ground state.
It is worth noting that the in-beam $\gamma$-ray spectra for $^{61}$Mn and $^{62}$Mn
obtained by Valiente-Dob\'{o}n {\it et al.} \cite{val2008} include transitions of 
energies 157 and 155 keV, respectively, supporting the present
assignment of the 157-keV $\gamma$-ray transition in Fig.\ \ref{fig:gamma_Cronly}
to the decay of $^{61}$Cr.    

\begin{table}[ht]
\caption{Energies and absolute intensities of delayed $\gamma$ 
rays assigned to the decay of $^{61}$Cr.  The initial and final 
states for those transitions placed in the proposed 
$^{61}$Mn level scheme of Fig.\ \ref{fig:Decayscheme} are also indicated.} 
\label{tab:gammas}
\begin{ruledtabular}
\begin{tabular}{rrrrr}
             &                    &  $E _{initial}$
               & $E_{final}$ & Coincident         \\
$E_{\gamma}$ & $I^{abs}_{\gamma}$ &  (keV) 
               & (keV)       & $\gamma$ rays (keV)\\
 
\hline
$ 157.2 \pm 0.5$ & $ 9 \pm 2$ &  157 &   0 &     \\
$ 354.8 \pm 0.4$ & $16 \pm 2$ & 1497 & 1142 & 535, 1142\\
$ 534.6 \pm 0.5$ & $ 5 \pm 1$ & 2032 & 1497 & 355\\
$1142.2 \pm 0.4$ & $21 \pm 2$ & 1142 &   0 &  355\\
$1497.3 \pm 0.5$ & $ 9 \pm 2$ & 1497 &   0 &     \\
$1860.8 \pm 0.4$ & $20 \pm 2$ & 1861 &   0 &     \\
$2378.2 \pm 0.4$ & $11 \pm 1$ & 2378 &   0 &     \\
\end{tabular}
\end{ruledtabular}
\end{table}

\begin{figure}[h]
\includegraphics[width=0.5\textwidth]{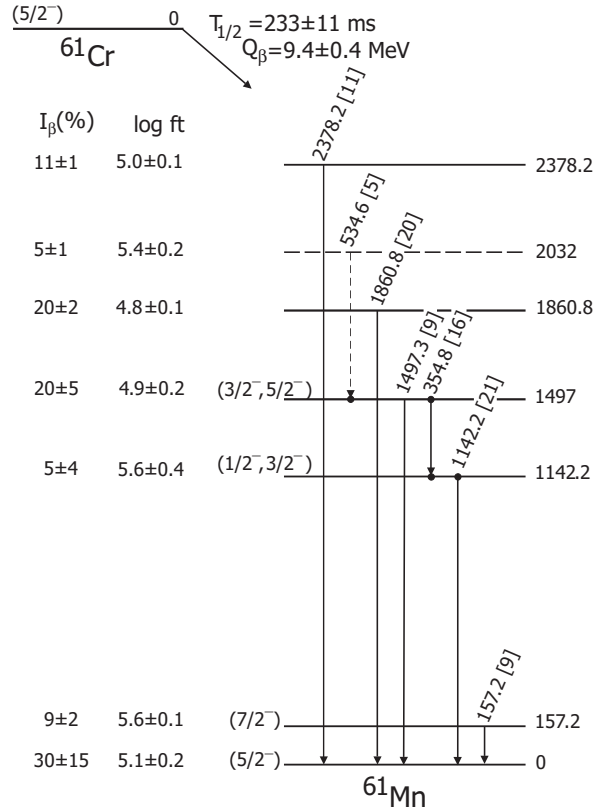}
\caption{Proposed $^{61}$Mn level scheme populated 
following the $\beta$ decay of $^{61}$Cr.  The number in brackets 
following a transition energy is the absolute 
$\gamma$-ray intensity.  The $Q$ value was taken 
from Ref.\ \cite{aud2003}.  Observed coincidences 
are represented by filled circles.  Absolute $\beta$-decay intensities 
and apparent log $ft$ values to each state in $^{61}$Mn are given 
on the left-hand side of the figure.  The dashed state at 2032 keV is 
tentatively placed as described in the text.}
\label{fig:Decayscheme}
\end{figure}

\begin{figure}[h]
\includegraphics[width=0.5\textwidth]{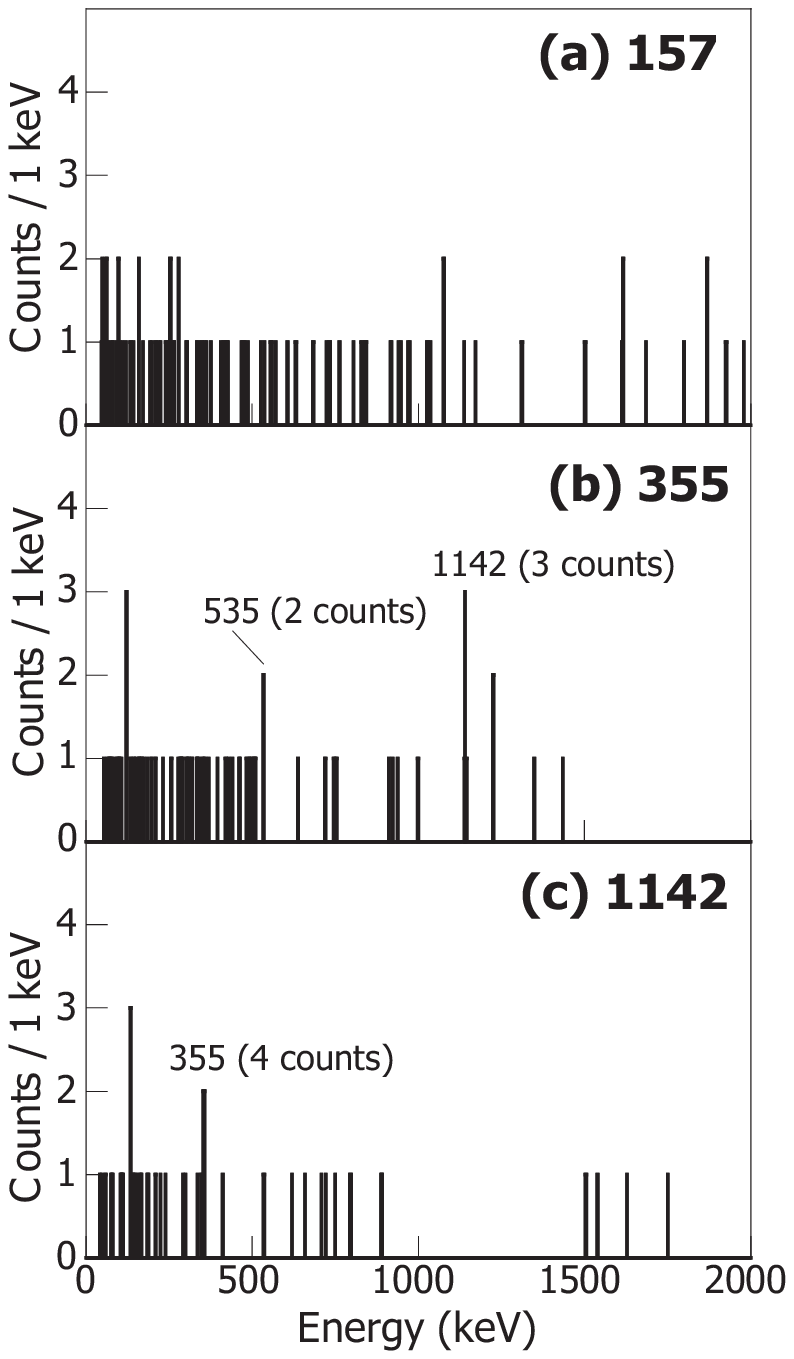}
\caption{$\gamma \gamma$ coincidence spectra 
for: (a) 157 keV; (b) 355 keV; and (c) 1142 keV $\gamma$ rays.  
The spikes observed below 200 keV in (b) and (c) represent 3 counts 
each in a single channel at low energy, and an absence of width 
uncharacteristic of a true $\gamma$-ray coincidence.}
\label{fig:gamma_gamma}
\end{figure}

The proposed decay scheme for levels in $^{61}$Mn populated 
following the $\beta$ decay of $^{61}$Cr is presented 
in Fig.\ \ref{fig:Decayscheme}.  The $\beta$-decay $Q$ 
value was taken from Ref.\ \cite{aud2003}.  
Absolute $\gamma$-ray intensities were deduced from 
the number of observed $^{61}$Cr $\gamma$ rays, 
the $\gamma$-ray peak efficiency, and the number 
of $^{61}$Cr implants correlated with $\beta$ decays, 
as derived from the fit of the decay curve in Fig.\ \ref{fig:halflife}.  
The two-$\gamma$ cascade involving the 
1142- and 355-keV transitions was confirmed by 
$\gamma \gamma$ coincidence relationships (see Fig.\ \ref{fig:gamma_gamma}).
However, the ordering of the two transitions is not 
uniquely determined.  The arrangement of 
Fig.\ \ref{fig:Decayscheme} was based on the 
absolute intensities of the 1142- and 355-keV transitions.  
No evidence was found for $\gamma$ rays in coincidence 
with the 157-keV ground-state transition within the statistical
uncertainty of the measurement.  Two 
counts in the 355-keV coincidence spectrum of Fig.\ \ref{fig:gamma_gamma}(b) 
suggested placement of the 535-keV line
in cascade from a higher-lying level at 2032 keV.  This tentative placement is 
represented in Fig.\ \ref{fig:Decayscheme} by a dashed line.

Apparent $\beta$-decay feeding to levels in $^{61}$Mn was deduced 
from the absolute $\gamma$-ray intensities.  The deduced 
values, along with apparent log $ft$ values, are also given
in Fig.\ \ref{fig:Decayscheme}.  
The observed $\beta$ branches all have apparent log $ft$ 
values between 4 and 6, consistent with allowed transitions.  
The 1142-keV level is deduced to have a small
$\beta$ branching with a large error, and may  not 
be directly populated by the $\beta$ decay of $^{61}$Cr.  

A ground-state spin and parity assignment of 
$5/2^{-}$ was adopted for $^{61}$Mn, by 
Runte {\it et al.} \cite{run1985}, based on the systematic trends
of ground state $J^{\pi }$ values for the less neutron-rich, 
odd-$A$ Mn isotopes.  The first excited state at 157 keV has
been tentatively assigned $J^{\pi } = 7/2^{-}$ from the in-beam
$\gamma$-ray results \cite{val2008}.    Allowed $\beta$ feeding
to the lowest two states would then limit the $J^{\pi }$ quantum numbers 
of the $^{61}$Cr ground state to values of $5/2^-$ or $7/2^-$.
Gaudefroy {\it et al.} \cite{gau2005} tentatively assigned $J^{\pi } = 5/2^-$
to the ground state of $^{61}$Cr on the basis of the 
observed $\beta$-decay properties of the progenitor, $^{61}$V, 
and such an assignment is adopted in Fig.\ \ref{fig:Decayscheme}.
It is likely that the 1142-keV state has low spin, either $J ^{\pi } = 1/2^{-}$ 
or  $J ^{\pi } = 3/2^{-}$.  The level at 1142 keV is apparently weakly fed 
by $\beta$ decay, was not identified in the yrast structure of $^{61}$Mn \cite{val2008},
and does not depopulate to the $7/2^-$ level at 157 keV within the 
statistical certainty of this measurement.       
The 1497-keV level is tentatively assigned a $J^{\pi }$ 
of $3/2^-$ or $5/2^-$ because of the apparent allowed 
$\beta$-decay branch and the favorable competition 
between the two depopulating $\gamma$ rays 
that suggests both transitions have M1
multipolarity. 
The proposed levels at excitation energies of 1861 and 
2378 keV are apparently fed by allowed $\beta$ decay as well, and can take
$J^{\pi }$ values in a range from $3/2^-$ to $7/2^-$.  

\section{Discussion}

The possible increase in collectivity inferred at low energy 
for the neutron-rich $_{24}$Cr and $_{26}$Fe isotopes has been associated
with the presence of the neutron $1g_{9/2}$ single-particle
orbital near the Fermi surface as $N = 40$ is approached.
As noted in the introduction, the systematic 
variation in $E(2^+_1)$ values (Fig.\ \ref{fig:systematics})
may suggest that deformation sets in already at $N=36$
for the Cr isotopes, while first evidence for such deformation effects 
in the Fe isotopes occurs at $^{64}$Fe, which has $N=38$.  Therefore,
it could be speculated that the low-energy structure of 
$^{61}$Mn, with $N=36$, might already exhibit features 
at low energy suggestive of a change in collectivity 
when compared to other odd-$A$ Mn isotopes nearer to stability.

\subsection{Odd-$A$ Mn Isotopes}

\begin{figure}[h]
\includegraphics[width=0.5\textwidth]{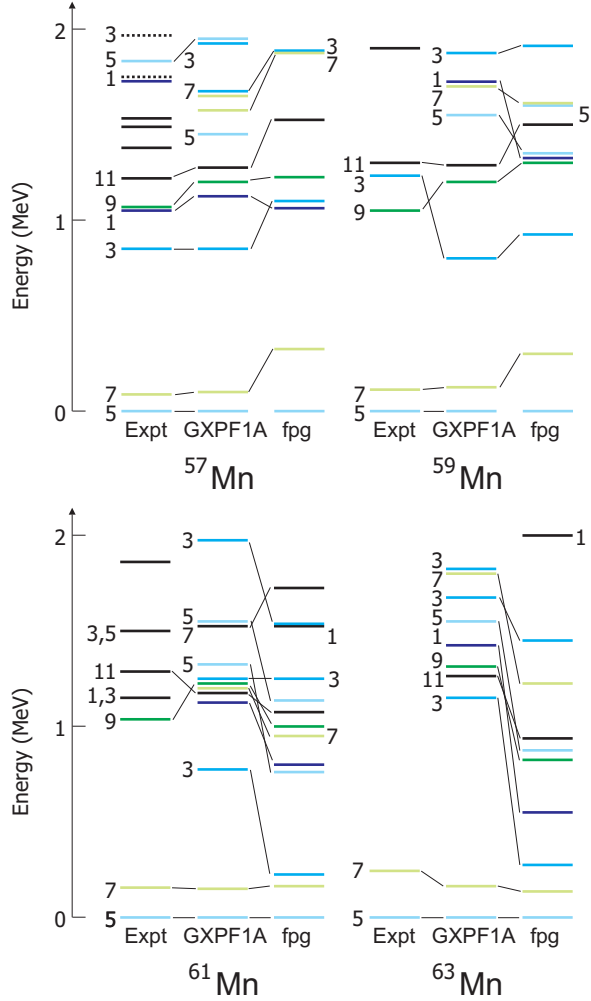}
\caption{(Color online) Systematic variation of the low-energy levels 
of the neutron-rich, odd-$A$ $_{25}$Mn isotopes, along
with the results of shell model calculations with the 
{\sc GXPF1A} and $fpg$ interactions. Only 
levels below 2~MeV are presented.  In addition,
the shell model results shown are limited to
the three lowest energy levels calculated for 
$J^{\pi } = 1/2^{-} - 7/2^{-}$, and the yrast levels for 
$J^{\pi } = 9/2^{-}$ and $11/2^{-}$.  Levels with assumed
negative parity are shown as solid lines, and those with
positive parity are presented as dotted lines.  Spins 
are given as $2I$ and, in nearly all cases, are 
tentative as discussed in the text.}
\label{fig:oddAMn}
\end{figure}

The systematics of the known 
energy levels of the odd-$A$ Mn isotopes 
with $A=57-63$ can be found in Fig.\ \ref{fig:oddAMn}.
The levels shown below 2~MeV for $^{57}$Mn are
taken from $\beta$ decay of $^{57}$Cr \cite{dav1978}, in-beam
$\gamma$-ray spectroscopy \cite{app2000}, and 
the (d,$^{3}$He) transfer 
reaction \cite{put1983}.  
$\beta$ decay apparently populates seven
excited states below 2~MeV, as well as the $^{57}$Mn ground state.  
The states directly fed by $\beta$ decay 
are all assumed to have negative parity, since
the $^{57}$Cr parent has ground state $J^{\pi } = 3/2^-$ quantum numbers.  
The negative parity yrast states observed in $^{57}$Mn 
were populated by a heavy-ion induced fusion-evaporation 
reaction, and resulted in tentative spin assignments 
to levels up to $J = 25/2$.  The (d,$^{3}$He) transfer
reaction provided unambiguous $J^{\pi}$  
assignments to the negative parity states below 2~MeV at 
84, 851, and 1837~keV, based on angular distribution
data.  Positive parity was deduced for the 
excited states with energies 1753 and 1965~keV.
Based on the spectroscopic strengths for proton
pickup \cite{put1983}, the first 
$7/2^-$, $1/2^+$, and $3/2^+$ levels in $^{57}$Mn  
were deduced to carry $\sim 40$\% of the
summed single-particle strength from shell
model expectations.

Valiente-Dob\'{o}n {\it et al.} \cite{val2008} 
studied neutron-rich Mn isotopes by in-beam $\gamma$-ray spectroscopy
following deep inelastic collisions of a $^{70}$Zn beam
on a $^{238}$U target.  
Negative-parity yrast states were identified 
up to $J = 15/2$ in $^{59}$Mn$_{34}$ and $J=11/2$ in $^{61}$Mn$_{36}$.  
Only a single transition, with energy 248~keV, was assigned 
to the structure of $^{63}$Mn$_{38}$.  
The non-yrast levels in $^{59,61}$Mn shown 
in Fig.\ \ref{fig:oddAMn} were identified
following $\beta$ decay.  The $\beta$ decay
of $^{59}$Cr to levels in $^{59}$Mn was most recently 
reported by Liddick {\it et al.} \cite{lid2005}.
The ground state spin and parity of the parent $^{59}$Cr, 
tentatively assigned as $J^{\pi } = 1/2^-$, restricts the range 
of states in $^{59}$Mn accessible by allowed $\beta$ decay
to those with $J^{\pi } = 1/2^-, 3/2^-$.
The ground state of $^{59}$Mn is known to have
$J^{\pi } = 5/2^-$ \cite{oin2001}.  The non-yrast
$J^{\pi }$ values for $^{59}$Mn were inferred from the 
$\beta$- and $\gamma$-decay patterns.   

The negative parity yrast structures below
2~MeV in odd-$A$ $^{57,59,61}$Mn exhibit
little variation as a function of neutron number.  These 
levels are consistent with shell model calculations
reported in Ref.\ \cite{val2008} employing the 
{\sc GXPF1A}, {\sc KB3G} and $fpg$ effective interactions.
The authors of Ref.\ \cite{val2008} do note, however,
that the proper ordering of the $9/2^-$ and $11/2^-$ 
levels at $\sim 1$~MeV is reproduced for all three
isotopes only by the $fpg$ interaction.  The experimental
energy gap between the ground state, with $J^{\pi } = 5/2^-$, and 
the $7/2^-$ first excited state exhibits a regular increase from
$^{57}$Mn (83~keV) to $^{65}$Mn (272~keV) \cite{gau2005}.  Both the 
{\sc GXPF1A} and $fpg$ shell model interactions reproduce the 
observed trends well.  

The shell model calculations in Ref.\ \cite{val2008}
were extended to include non-yrast states below 2~MeV.  
The calculations were performed with 
the code {\sc ANTOINE} \cite{cau1989} and the
{\sc GXPF1A} \cite{hon2005} and $fpg$ \cite{sor2002} 
effective interactions.  The full $fp$ model space 
was utilized for the calculations with the 
{\sc GXPF1A} interaction.  The valence space
for the calculations with the 
cross shell interaction $fpg$ was limited to 
2 neutron excitations from the $fp$ shell to the 
$1g_{9/2}$ orbital.  As compared to Ref.\ \cite{val2008}, the 
truncation of the basis 
was necessary to reach convergence for the higher
density of states at lower spin for the most neutron-rich
Mn isotopes considered here.  However, in
those cases where the calculations were completed 
with both 2 and 6 neutron excitations allowed from the $fp$
shell, similar results were obtained. 

The low density of levels below 1~MeV is persistent
in $^{57,59,61}$Mn, and follows the shell model
results with the {\sc GXPF1A} effective interaction well.  
The only excited state in this
energy range is the $7/2^-$ level, which is expected to 
carry a reduced
$1f_{7/2}$ single-particle component with the addition of neutrons \cite{put1983}.
The low-spin level density below 1~MeV is shown to
increase at $^{61}$Mn in the shell model results with
the $fpg$ interaction.  Here, the influence of the $1g_{9/2}$
neutron orbital becomes apparent at $N=36$.  However, this
is not borne out in the observed level structure.  
The regular behavior of the low-energy levels of the 
odd-$A$ $_{25}$Mn isotopes through $N=36$ follows 
the similar trend observed in the even-even $_{26}$Fe
isotopes, where the possible onset of collectivity at low energy
is not evident until $N=38$ is reached.

\subsection{Even-$A$ Mn Isotopes}

Although the present experiment did not report
any new information on even-$A$ Mn isotopes, for the
sake of completeness we include them in the discussion.
Only a few excited states in the odd-odd $^{60,62}$Mn
are established, based on $\beta$-decay
studies.  Two $\beta$-decaying states are known
in both isotopes, but the location and ordering of the 
proposed $1^+$ and $4^+$ states is not established
in $^{62}$Mn$_{37}$ \cite{gau2005}.    
Good agreement was noted between shell model
results with the {\sc GXPF1} interaction and both the 
low-energy structure and $\beta$-decay properties of 
$^{60}$Mn$_{35}$ \cite{lid2006}.  The low-energy 
structure of $^{62}$Mn built on the $1^+$ $\beta$-decaying state
shows marked similarity to that in $^{60}$Mn,
albeit the excited $2^+$ and $1^+$ states are both shifted
$\sim 100$~keV lower in energy with the addition of two neutrons.  
Although Valiente-Dob\'{o}n {\it et al.}
reported a number of in-beam $\gamma$ rays associated with 
the depopulation of yrast levels in both $^{60}$Mn and $^{62}$Mn,
no level structures were proposed due to the lack of 
coincidence data in the literature \cite{val2008}.  A key feature
yet to be identified in the neutron-rich, odd-odd
Mn isotopes is the location of the negative-parity levels 
that could signify the presence of 
the intruding $1g_{9/2}$ neutron orbital.
Again, the systematic variation of the (few) excited levels 
established in the odd-odd Mn isotopes through $N=37$
does not suggest a sudden onset of collectivity, in line with the 
trend established for the yrast states of even-even $_{26}$Fe
isotopes.

\section{Summary}
The $\beta$ decay of $^{61}$Cr has been studied to 
extract details on non-yrast excited states in the daughter
nucleus $^{61}$Mn.  A more accurate half-life of $233 \pm 11$~ms 
was deduced for the 
ground-state of $^{61}$Cr, consistent with the 
previous  measurement by Sorlin {\it et al.} 
\cite{sor2000}.  The low-energy level scheme for $^{61}$Mn,
deduced following $\beta$ decay of $^{61}$Cr, features 
five new excited states above a 1-MeV excitation energy.  However,
the low-energy structure below 1~MeV does not resemble that expected
from the shell model results that consider the 1$g_{9/2}$ orbital. 
The low density of states below 1~MeV is a consistent
feature observed in the odd-$A$ Mn isotopes with $A=57,59,61$,
and follows the results of shell model calculations in
the $fp$ model space, without a need to invoke neutron
excitations into the $1g_{9/2}$ orbital.
There is also little variation in the yrast 
structures of these odd-$A$ Mn isotopes up to $J=15/2$, as reported by 
Valiente-Dob\'{o}n {\it et al.} \cite{val2008}.
No compelling evidence was found for a possible onset of 
collectivity in the low-energy structures of the 
Mn isotopes through $N=37$.  This observation is consistent
with the behavior of the even-even, $Z+1$ $_{26}$Fe isotopes, but
not with that of the even-even $Z-1$ $_{24}$Cr isotopes, where 
possible onset of collectivity at low energy has been suggested 
from the E($2^+_1$) energy at $N=36$. 
Scant data are available for both yrast and non-yrast levels
in the $_{25}$Mn isotopes beyond $N=36$.  Such data will be
critical to evaluate the role of the neutron $1g_{9/2}$ orbital
in defining the structural properties of the neutron-rich Mn isotopes.      

\begin{acknowledgments}
The authors thank the NSCL operations staff for providing the 
primary and secondary beams for this experiment and 
the NSCL $\gamma$ group for assistance in setting up the 
Ge detectors from SeGA. 
This work was supported in part by the National Science Foundation 
Grant No. PHY-06-06007, the U.S.\ Department of Energy, Office
of Nuclear Physics, under contracts DE-AC02-06CH11357 and DE-FG02-94ER40834,
and the Polish Academy of Sciences grant 1PO3B 059 29.  
HLC acknowledges support from the Natural Science and 
Engineering Research Council (NSERC) of Canada.
\end{acknowledgments}

\end{document}